\documentstyle[amssymb,twocolumn,aps]{revtex}

\input psfig.sty
\begin{document}
\title{CURRENT-INDUCED SUPERCONDUCTOR-INSULATOR TRANSITION IN GRANULAR HIGH-T$_{c}$
SUPERCONDUCTORS}
\author{Y. Kopelevich*, C. A. M. dos Santos**, S. Moehlecke*, and A. J. S. Machado**}
\address{(*) Instituto de F\'{i}sica ''Gleb Wataghin'', Universidade Estadual de
Campinas, Unicamp 13083-970, Campinas, S\~{a}o Paulo, Brasil\\
(**) Departamento de Engenharia de Materiais, FAENQUIL, 12600-000, Lorena, S%
\~{a}o Paulo, Brasil}
\maketitle

\section{INTRODUCTION}

The occurrence of either superconducting or insulating state in a
zero-temperature limit (T $\rightarrow $ 0) under variation of system
parameters such as, for instance, microscopic disorder in homogeneous thin
films \cite{Haviland,MPAFisherPRL1990,MCCha,YLiu,Markovic} or charging
energy/Josephson energy ratio in granular superconductors \cite
{Geerligs,Chen}, is one of the fundamental problems in the condensed matter
physics which continuously attracts an intense research interest. It has
also been shown both theoretically \cite{FisherPRL1990a}and experimentally
that the superconductor-insulator transition in two-dimensional (2D)
superconductors can be tuned by applied magnetic field. The field-tuned
superconductor-insulator transition has been measured in amorphous \cite
{Markovic,Hebard,Paalanen,Yazdani} and granular \cite{Kobayashi} thin films,
fabricated 2D Josephson-junction arrays \cite{Zant,Chen1995}, and bulk
layered quasi-2D superconductors such as high-T$_{c}$ cuprates \cite
{Seidler,Tanda}.

On the other hand, in granular superconductors both applied magnetic field
and electrical current affect the Josephson coupling between grains \cite
{Tinkham,Belevtsev,Gerber}. Besides, a zero-temperature current-driven
dynamic transition from a vortex glass to a homogeneous flow of the vortex
matter is expected to occur in disordered Josephson junction arrays \cite
{Dominguez}, suggesting the intriguing possibility of a current-induced
superconductor-insulator transition, analogous to the field-tuned transition 
\cite{Carlos}.

In this paper we report a systematic study of electrical current effects on
the superconducting properties of granular high-T$_{c}$ superconductors. The
here presented results demonstrate the occurrence of
superconductor-insulator quantum phase transition driven by the applied
electrical current, and provide evidence that the dynamics of Josephson
intergranular vortices plays a crucial role in this phenomenon.

\section{SAMPLES AND EXPERIMENTAL DETAILS}

Polycrystalline single phase Y$_{1-x}$Pr$_{x}$Ba$_{2}$Cu$_{3}$O$_{7-\delta }$
samples were prepared using a solid state reaction method with a route
similar as described in \cite{Maple} and characterized by means of x-ray
powder diffractometry, and optical and scanning electron microscopy. The
analysis showed that the samples are granular materials consisting of grains
with an average size d $\sim $ 5 $\mu $m. The superconducting transition was
measured both resistively and with a SQUID magnetometer MPMS5 (Quantum
Design). Electrical transport dc measurements in applied magnetic field H $%
\leq $\ 100 Oe, produced by a copper solenoid, were performed using standard
four-probe technique with low-resistance ($<$ 1 $\Omega $) sputtered gold
contacts. No heating effects due to current were observed for I \ $\leq $
100 mA, our largest measuring current. In order to eliminate thermoelectric
effects, the measurements were performed inverting the applied current.

Below we present the results of measurements performed on Y$_{1-x}$Pr$_{x}$Ba%
$_{2}$Cu$_{3}$O$_{7-\delta }$ sample with x = 0.45 close to the critical Pr
concentration x$_{c}$ $\approx $ 0.57 above which superconductivity has not
been detected \cite{Maple}. The sample superconducting transition
temperature (onset) T$_{c0}$ = 33 K, the resistivity at the superconducting
transition $\rho _{N}$(T$_{c0}$) = 20 m$\Omega $cm, and dimensions l x w x t
= 12.6 x 1.94 x 1.24 mm$^{3}$.

\section{RESULTS AND DISCUSSION}

\begin{figure}[t]
\centerline{\psfig{file=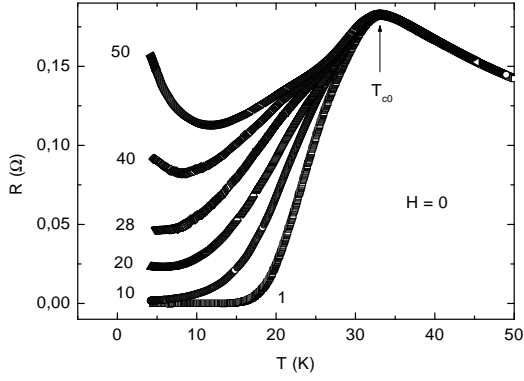,width=85mm}}
\caption{Resistance R(T) = V(T)/I obtained at various measuring currents
(here and in Figs. \ref{studfig2} - \ref{studfig4}, numbers at the curves correspond to the applied
current in mA) and at no applied magnetic field.}
\label{studfig1}
\end{figure}

\begin{figure}[t]
\centerline{\psfig{file=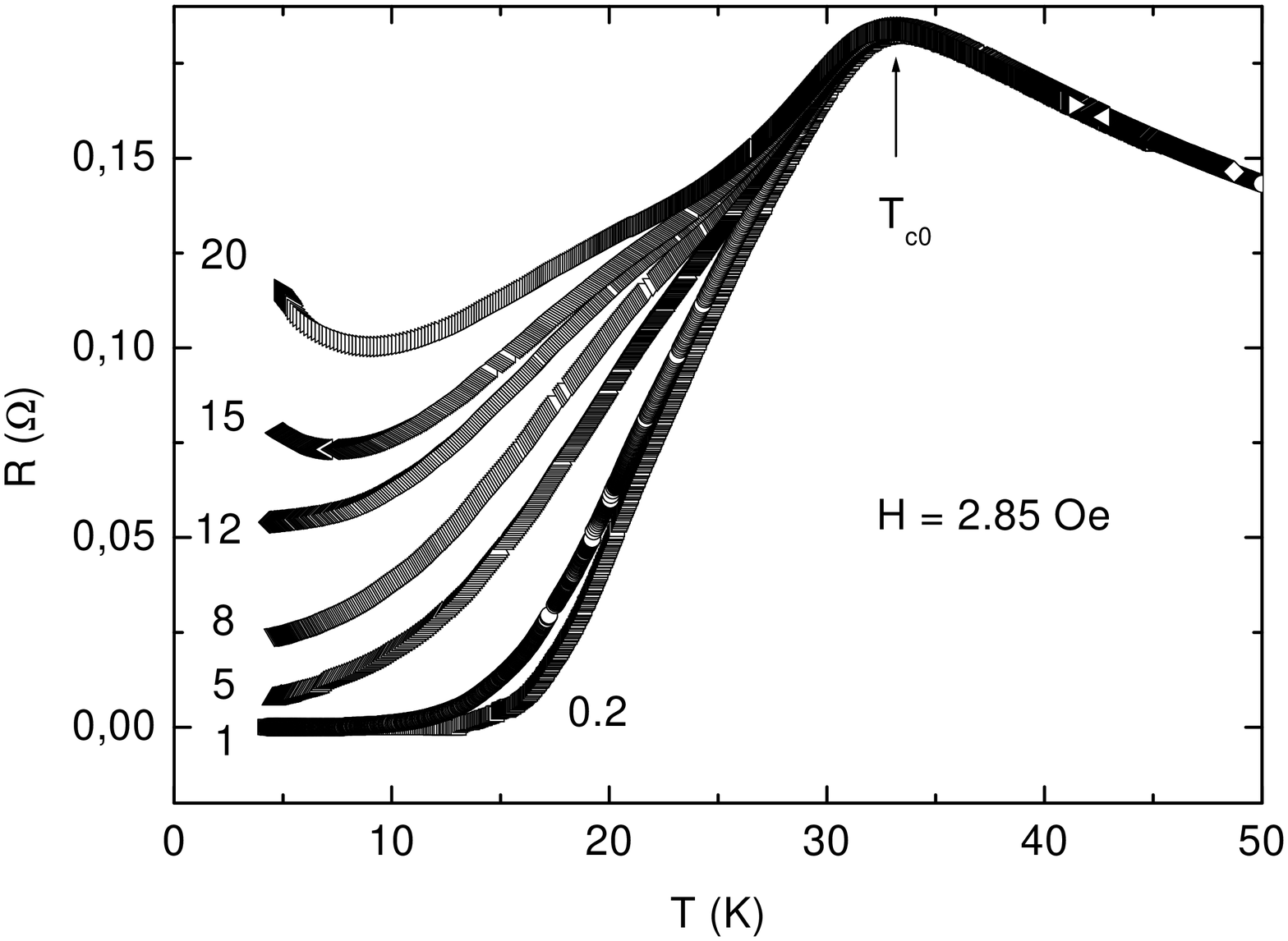,width=85mm}}
\caption{Resistance R(T) = V(T)/I obtained at various measuring currents and
applied magnetic field H = 2.85 Oe.}
\label{studfig2}
\end{figure}

\begin{figure}[t]
\centerline{\psfig{file=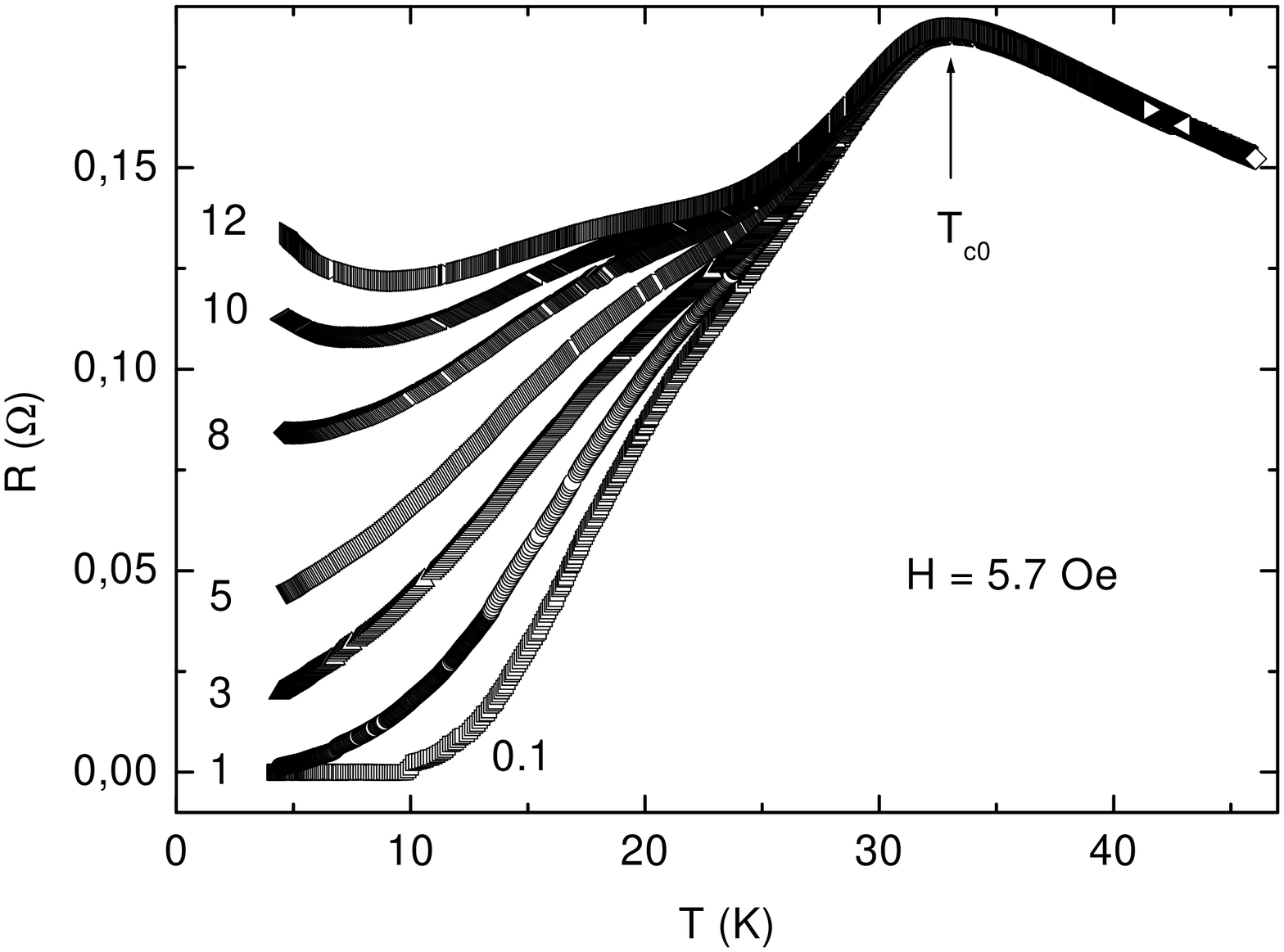,width=85mm}}
\caption{Resistance R(T) = V(T)/I obtained at various measuring currents and
applied magnetic field H = 5.7 Oe.}
\label{studfig3}
\end{figure}

\begin{figure}[t]
\centerline{\psfig{file=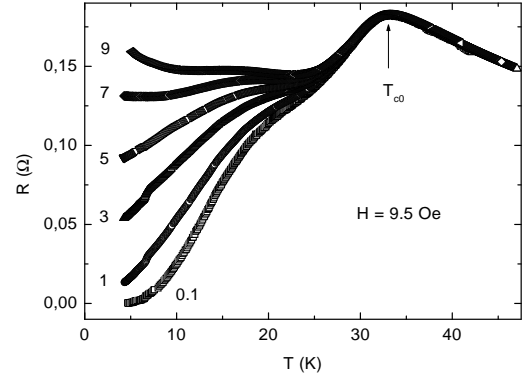,width=85mm}}
\caption{Resistance R(T) = V(T)/I obtained at various measuring currents and
applied magnetic field H = 9.5 Oe.}
\label{studfig4}
\end{figure}

Figures \ref{studfig1} -- \ref{studfig4} present the resistance vs. temperature data obtained in a
vicinity of T$_{c0}$ = 33 K for various applied currents and magnetic
fields. As Figs. \ref{studfig1} -- \ref{studfig4} illustrate, the superconducting transition
temperature onset T$_{c0}$ is both current- and field-independent. On the
other hand, the zero-resistance superconducting state is destroyed by the
application of both the electrical current and magnetic field.

The superconducting order parameter $\Psi =\Psi _{0}e^{i\phi }$ has two
components; a magnitude $\Psi _{0}$ and a phase $\phi $. Because T$_{c0}$
(and hence $\Psi _{0}$) remains unchanged, the low-temperature finite
resistance in our sample originates from thermal and/or quantum fluctuations
in the phase locking. In the context of granular superconductors, T$_{c0}$
can be identified with the transition temperature of individual grains, and
the phase fluctuations -- with fluctuating Josephson currents between
grains. In the absence of external field, the critical current is that
corresponding to a vortex creation and its motion, neglecting pinning.
However, in granular superconductors with a very weak coupling between
grains, the Earth's magnetic field H$_{E}$ $\sim $ 0.5 Oe (which has not
been shielded in the experiments) can easily penetrate the sample. Then, the
critical

\noindent
current will be determined by the pinning of Josephson
intergranular vortices originating, for example, from an inhomogeneity of
the Josephson junction coupling strength.

\begin{figure}[t]
\centerline{\psfig{file=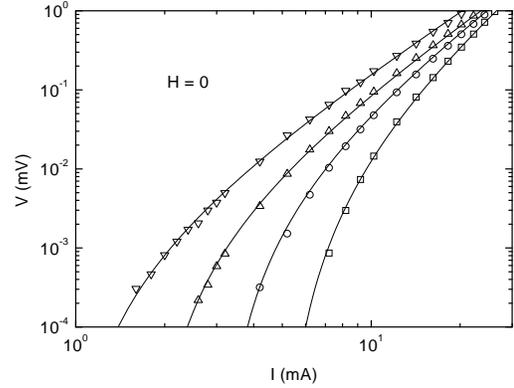,width=85mm}}
\caption{ Low-current portions of I-V isotherms measured at T = 4.6 K ($\Box $%
), T = 8 K (o), T= 11 K ($\triangle $), and T = 13 K ($\nabla $). Solid
lines are obtained from Eq. (1) with the fitting parameters c = 0.0001 mV/mA$%
^{n}$, I$_{th}$ = 5 mA, n = 3 (T = 4.6 K); c = 0.00025 mv/mA$^{n}$, I$_{th}$
= 3.1 mA, n = 2.7 (T = 8 K); c = 0.00039 mv/mA$^{n}$, I$_{th}$ = 1.8 mA, n = 2.55
(T = 8 K); ); c = 0.0007 mv/mA$^{n}$, I$_{th}$ = 0.94 mA, n = 2.45 (T = 13
K).}
\label{studfig5}
\end{figure}

Shown in Fig. \ref{studfig5} are low-current portions of current-voltage I-V
characteristics measured at low temperatures and at no applied field. As can
be seen from Fig. \ref{studfig5}, the V(I) dependencies can be very well described by the
equation

\begin{equation}
\stackrel{}{V=c(T,H)(I-I_{th}(T,H))^{n(T,H)}},  \label{Eq.1}
\end{equation}

where I$_{th}$(T,H) is the threshold current. Thus, at T = 4. 6 K, I$_{th}$
= 5 mA and the corresponding current density ( I$_{th}$/cross section of the
sample) j$_{th}$ = 0.21 A/cm$^{2}$.

Within the framework of effective Josephson medium theory \cite
{Clem,Sonin,Lobb} one can estimate the lower critical field H$_{c1J}$ $\sim $
8$\pi ^{2}$j$_{c0}\lambda _{L}$/c $\sim $ 0.2 mOe, taking the maximum
Josephson current density j$_{c0}$ $\sim $ 10j$_{th}$ \cite{Clem,Abraham}
and the typical value of the intragranular London penetration depth $\lambda
_{L}$ $\sim $ 0.1 $\mu $m ($<$$<$ d). The obtained value of H$_{c1J}$ is
much smaller than the Earth's magnetic field, indeed, i. e. our sample is
deeply in the mixed state.

Below the threshold current I$_{th}$(H,T) the vortex motion is strongly
suppressed or zero. The I-V characteristics described by the Eq. (\ref{Eq.1}%
) are expected in the regime where the interaction between vortices and the
pinning potential dominates the vortex-vortex interaction \cite
{Dominguez,Kes,Higgins,Enomoto}.

\begin{figure}[t]
\centerline{\psfig{file=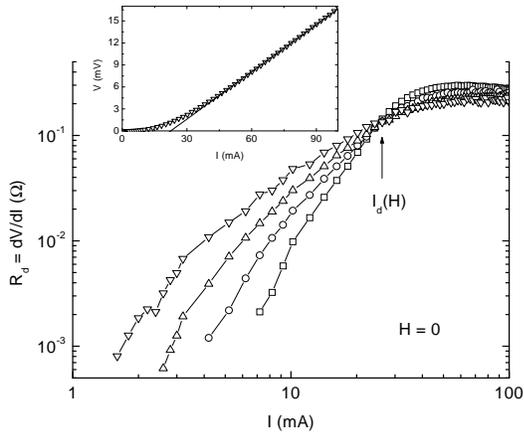,width=85mm}}
\caption{Differential resistance R$_{d}$ = dV/dI obtained at no applied field
and T = 4.6 K ($\Box $), T = 8 K (o), T= 11 K ($\triangle $), and T = 13 K ($%
\nabla $). Inset shows V(I) measured at T = 13 K; the solid line is obtained
from the Eq. (1) with R$_{ff}$ = 0.215 $\Omega $ and I$_{cf}$ = 22.5 mA.}
\label{studfig6}
\end{figure}

As the applied current increases, the V(I) curves approach a linear regime
of flux flow where the differential resistance R$_{d}$ = dV/dI is
current-independent, see Fig. \ref{studfig6}. Inset in Fig. \ref{studfig6}
 exemplifies V(I) measured at T = 13 K and demonstrates that at large enough
 currents the V(I) can be
fitted by the equation

\begin{equation}
V(I)=R_{ff}(I-I_{cf}).  \label{Eq.2}
\end{equation}

Here R$_{ff}$ is the flux-flow resistance and I$_{cf}$ is the so-called
dynamical critical current. In this high-current regime vortices move
coherently, only weakly interacting with the pinning potential.

As can also be seen in Fig. \ref{studfig6}, there exists a crossing of R$_{d}$(I) curves
occurring at I = I$_{d}$. The measurements performed at various applied
fields revealed that I$_{d}$(H) is a decreasing function of field. Such a
crossing has also been measured in Mo$_{77}$Ge$_{23}$ films \cite{Ephorn}
and reproduced in numerical simulations \cite{Ryu} which show that the
pinned fraction of vortices rapidly decreases for I $>$ I$_{d}$(H).

\begin{figure}[t]
\centerline{\psfig{file=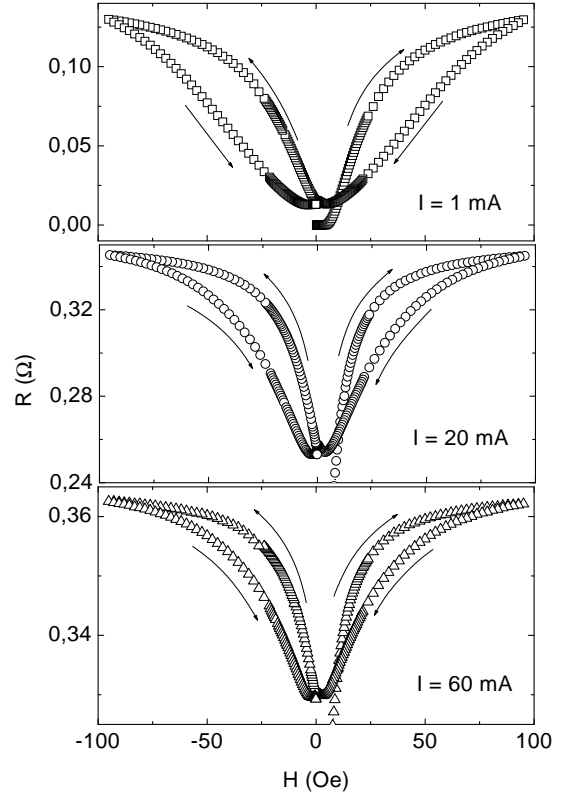,width=85mm}}
\caption{R(H) magnetoresistance hysteresis loops measured at T = 4.2 K and
applied currents I = 1 mA, I = 20 mA, and I = 60 mA, demonstrating that the
sample resistance is governed by the motion of intergranular Josephson
vortices; arrows indicate the magnetic field direction in the measurements.}
\label{studfig7}
\end{figure}

Figure \ref{studfig7} presents typical resistance hysteresis loops R(H) measured at T =
4.2 K for various currents. These measurements were performed after cooling
the sample from T $>$ T$_{c0}$ to the target temperature in a zero applied
field. As shown in Fig. \ref{studfig7}, the resistance R corresponding to the increasing $%
|$H$|$ branch is larger than that corresponding to the decreasing $|$H$|$
branch resulting in ``clockwise'' R(H) hysteresis loops. Previous studies 
\cite{Yakov,Ji,Li} showed that the ``clockwise'' R(H) hysteresis loops are
essentially related to the granular sample structure where the resistance is
determined by the motion of Josephson intergranular vortices. In increasing
magnetic field larger than the first critical field of the grains (H $>$ H$%
_{c1g}$), the pinning prevents Abrikosov vortices from entering the grains,
and flux lines (Josephson vortices) are concentrated between grains leading
to the intergrain field higher than the external applied field. When the
applied field is decreased, the pinning prevents Abrikosov vortices being
expelled from the grains resulting in a lower density of Josephson
intergrain vortices as compared to that in the increasing field. Thus, the
data of Fig. \ref{studfig7} provide an unambiguous evidence that the resistance in our
sample is governed by the motion of intergranular Josephson vortices in both
low- and high-current limits.

\begin{figure}[t]
\centerline{\psfig{file=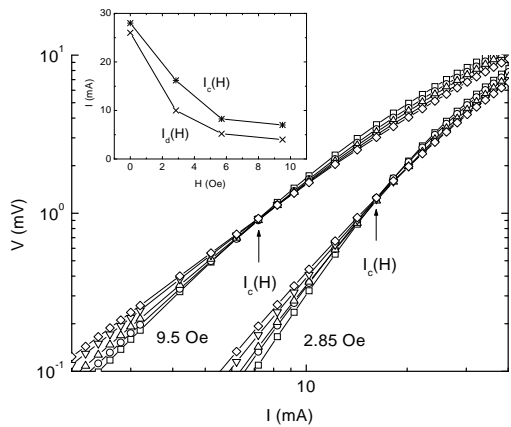,width=85mm}}
\caption{Current-voltage I-V characteristics measured in applied fields H =
2.85 Oe and H = 9.5 Oe at temperatures 4.6 K ($\Box $), 5.6 K (o), 6.7 K ($%
\triangle $), 8 K ($\nabla $), and 9 K ($\diamondsuit $). Inset shows I$_{d}$%
(H) and I$_{c}$(H) magnetic field dependencies.}
\label{studfig8}
\end{figure}

We note further that the resistance R(T) = V(T)/I in the low-temperature
limit reveals a crossover from the metallic-like (dR/dT $>$ 0) to the
insulating-like (dR/dT $<$ 0) behavior as the applied current is increased,
see Figs. \ref{studfig1} - \ref{studfig4}. The results presented in Figs. \ref{studfig1} - \ref{studfig4} indicate also the
existence of a field-dependent current I$_{c}$(H), separating the metallic-like
(I $<$ I$_{c}$(H)) and insulating-like (I $>$ I$_{c}$(H)) resistance curves,
at which the resistance is temperature-independent. The occurrence of the
temperature-independent resistance R$_{c}$ = (V/I)$|_{I_{c}}$ can clearly be
seen in Fig. \ref{studfig8} where a crossing of current-voltage I-V isotherms at I$_{c}$%
(H) measured for two applied fields is shown. The inset in Fig. \ref{studfig8} depicts I$%
_{c}$(H) and I$_{d}$(H) which illustrates that for a given field I$_{c}$(H) $%
>$ I$_{d}$(H). In other words, the crossing of I-V curves takes place in the
regime where the interaction between Josephson vortices dominates that with
the pinning potential.

These results resemble very much the resistance behavior in a vicinity of
the magnetic-field-tuned superconductor-insulator transition \cite
{Hebard,Paalanen,Yazdani,Kobayashi,Zant,Chen1995,Seidler,Tanda}. According
to the theory \cite{FisherPRL1990a}, at low enough magnetic fields vortices
are localized by the quenched disorder, leading to the zero-resistance (R =
0) superconducting state. With increasing vortex density and above some
critical field H$_{c}$ vortices delocalize and Bose condense, leading to the
insulating state.

\begin{figure}[t]
\centerline{\psfig{file=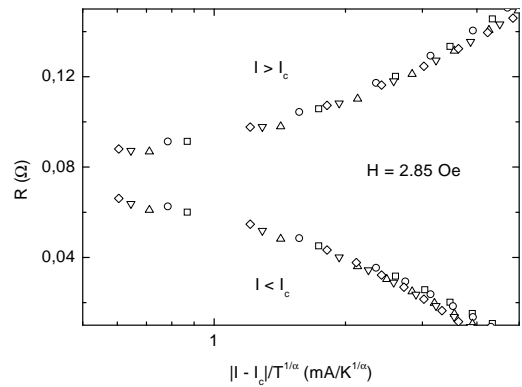,width=85mm}}
\caption{Resistance R = V/I as a function of the scaling variable obtained
for applied field H = 2.85 Oe (I$_{c}$ = 16.2 mA, $\alpha $ = 1.84); T = 4.6
K ($\Box $), T = 5.6 K (o), T = 6.7 K ($\triangle $), T = 8 K ($\nabla $).}
\label{studfig9}
\end{figure}

\begin{figure}[t]
\centerline{\psfig{file=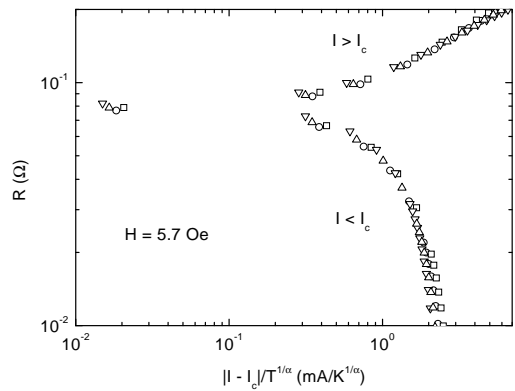,width=85mm}}
\caption{Resistance R = V/I as a function of the scaling variable obtained
for applied field H = 5.7 Oe (I$_{c}$ = 8.25 mA, $\alpha $ = 1.72); T = 4.6
K ($\Box $), T = 5.6 K (o), T = 6.7 K ($\triangle $), T = 8 K ($\nabla $).}
\label{studfig10}
\end{figure}

\begin{figure}[t]
\centerline{\psfig{file=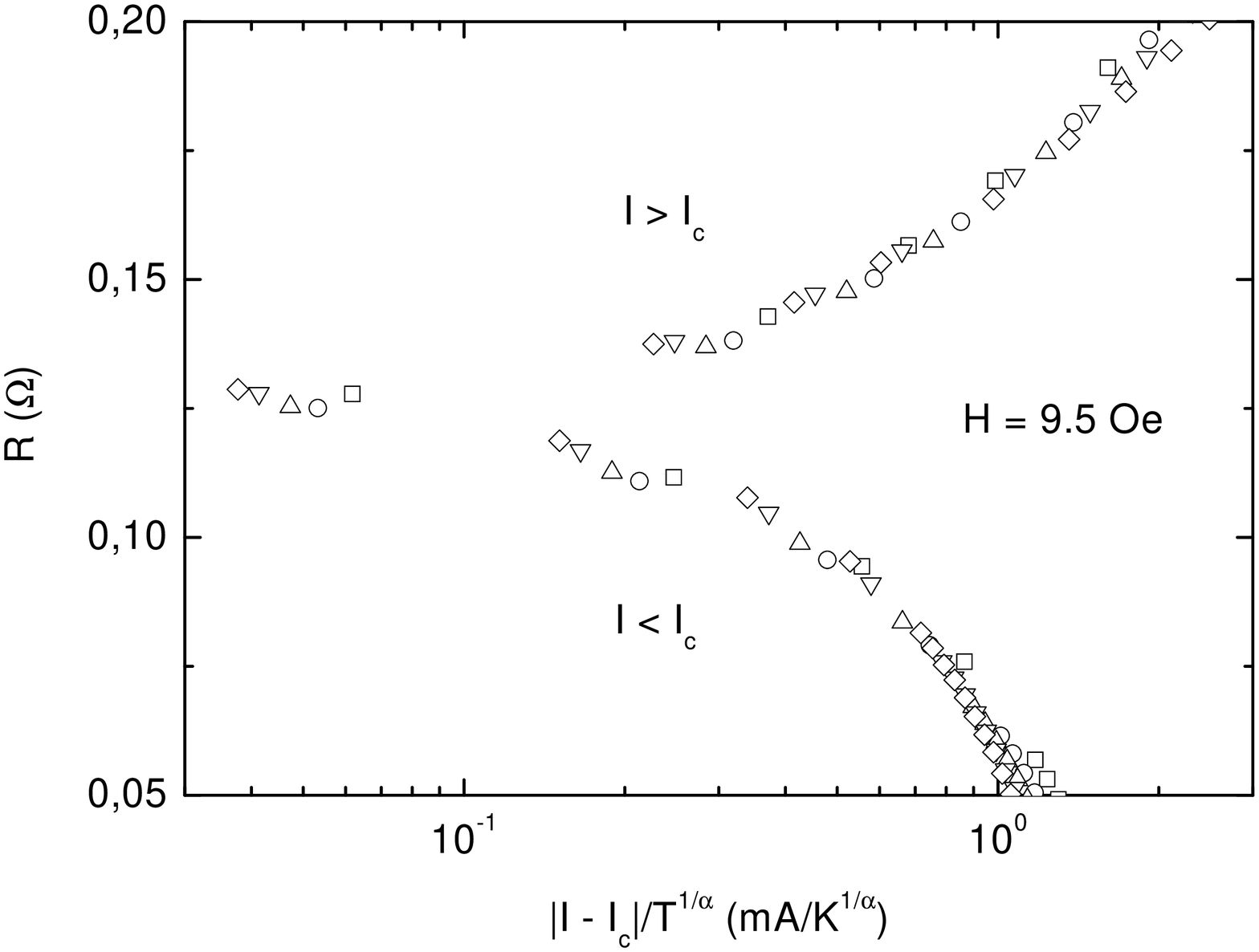,width=85mm}}
\caption{ Resistance R = V/I as a function of the scaling variable obtained
for applied field H = 9.5 Oe (I$_{c}$ = 7 mA, $\alpha $ = 1.3); T = 4.7 K ($%
\Box $), T = 5.7 K (o), T = 6.7 K ($\triangle $), T = 8 K ($\nabla $), T = 9
K ($\diamondsuit $).}
\label{studfig11}
\end{figure}

The theory predicts a finite-temperature scaling law \cite
{MPAFisherPRL1990,MCCha,FisherPRL1990a} which allows for experimental
testing of the zero-temperature superconductor-insulator transition. It is
assumed in this analysis that a sample is superconductor or insulator at T =
0 when dR/dT $>$ 0 or dR/dT $<$ 0, respectively. Besides, because the
magnitude of the superconducting order parameter $\Psi _{0}$ (T$_{c0}$)
remains unchanged on each grain as the intergranular resistance R(T, H, I)
undergoes the transition from metallic-like to insulating-like behavior, the
hard-core boson model \cite{MPAFisherPRL1990,MCCha,FisherPRL1990a} should be
a good approximation in our analysis of the superconductor-insulator
transition driven by the applied current. The resistance in the critical
regime of the quantum transition is given by the equation

\begin{equation}
R(\delta ,T)=R_{c}f(|\delta |/T^{1/z\nu }),  \label{Eq.3}
\end{equation}

where R$_{c}$ is the resistance at the transition, f($|\delta $$|$/T$%
^{1/z\nu }$) is the scaling function such that f(0) = 1, z and $\nu $ are
critical exponents, and $\delta $ is the deviation of a variable parameter
from its critical value. Assuming $\delta $ = I - I$_{c}$ ($\delta $ = H - H$%
_{c}$ in the case of a field-tuned transition) we plotted in Figs. \ref{studfig9} - \ref{studfig11},
R = V/I vs. $|\delta $$|$/T$^{1/\alpha }$ for three applied fields. In each
case, the exponent $\alpha $ was obtained from log-log plots of (dR/dI)$%
|_{I_{c}}$ vs. T$^{-1}$. The expected collapse of the resistance data onto
two branches distinguishing the I $<$ I$_{c}$ from the I $>$ I$_{c}$ data is
very clear in Figs. \ref{studfig9} - \ref{studfig11}.

Although, in the absence of a proper theory, the results presented in Figs.
\ref{studfig9} - \ref{studfig11} should be considered as empirical ones, the very good scaling fit
strongly suggests the occurrence of a zero-temperature current-induced
superconductor-insulator transition. Notably, the here found exponents $%
\alpha $ agree nicely with most of the data obtained in the scaling analysis
of the field-tuned transition \cite
{Markovic,Hebard,Paalanen,Yazdani,Kobayashi,Zant,Chen1995,Seidler,Tanda}.

\section{CONCLUDING REMARKS}

In conclusion, we would like to emphasize that the here studied
superconductor-insulator transition driven by the electrical current is
essentially related to the dynamics of Josephson intergranular vortices.
Actually, as the above results demonstrate, the crossing of I-V
characteristics takes place in the flux flow regime. We stress that the
crossover current I$_{c}$(H) is smaller than the maximum supercurrent I$_{0}$
= (2e/$\hbar $)E$_{J}$ (E$_{J}$ is the Josephson coupling energy) that the
Josephson junctions can support. In other words, both tunneling Cooper pairs
and moving Josephson vortices are present at the I$_{c}$(H) which play a
dual role as discussed in the context of the field-tuned transition \cite
{FisherPRL1990a}.

A self-consistent analysis would require the existence of zero-temperature
``depinning'' transition in the Josephson vortex matter driven by the
applied current. The occurrence of such a transition in disordered Josephson
junction arrays has theoretically been predicted, indeed \cite{Dominguez}.

All these indicate that the current-induced superconductor-insulator
transition can be considered as the dynamical counterpart of the
magnetic-field-tuned transition.

Finally, we point out that similar results were obtained on Y$_{1-x}$Pr$_{x}$%
Ba$_{2}$Cu$_{3}$O$_{7-\delta }$ (x = 0.5) and Bi$_{2}$Sr$_{2}$Ca$_{1-x}$Pr$%
_{x}$Cu$_{2}$O$_{8+\delta }$ (x = 0.54) granular superconductors.

\section{ACKNOWLEDGEMENTS}

The authors would like to thank Enzo Granato for the fruitful discussions.
This study was supported by FAPESP and CNPq Brazilian science agencies.

\end{document}